\documentclass{ws-mpla}
\usepackage{graphicx}
\usepackage{amsmath}
\begin{document}

\title{\bf EXPLORING THE COSMIC MICROWAVE BACKGROUND AS A COMPOSITION OF SIGNALS WITH KOLMOGOROV ANALYSIS}

\author{S.MIRZOYAN and E.POGHOSIAN}
\address{Yerevan Physics Institute and Yerevan State University, Yerevan, Armenia}

\maketitle

\pub{Received (\today)}{Revised (\today)}

\begin{abstract}
The problem of separation of different signals in the Cosmic Microwave Background (CMB) radiation using the difference in their statistics is analyzed. Considering  samples of sequences which model the CMB as a superposition of signals, we show how the Kolmogorov stochasticity parameter acts as a relevant descriptor, either qualitatively or quantitatively, to distinguish the statistical properties of the cosmological and secondary signals.  
\end{abstract}

\section{Introduction}

In the study of the Cosmic Microwave Background radiation, one of the basic information sources on the early Universe, one deals with a principal problem of separation of the cosmological signal from perturbations of various nature \cite{H}.The CMB reflects not only effects occurring at the early or large scale Universe, the Sachs-Wolf and Sunyaev-Zeldovich effects, but also more local perturbations of the Galactic, interplanetary, and up to instrumental origin.

Various descriptors are applied to separate the CMB cosmological signal from a secondary signal of particular nature. In the present paper we will apply the Kolmogorov stochasticity parameter \cite{Kolm,Arnold} to reveal the difference in the statistical properties of contributions irrelevant to the main CMB signal. As was already shown, the Kolmogorov's parameter is an informative descriptor when applied to the CMB temperature sequences, namely, it enables one to locate regions of different randomness in the sky maps \cite{GK_KSP1,GK_KSP2}. For example, the Cold Spot, the non-Gaussian anomaly in the Southern sky, not only shows enhanced value of Kolmogorov's parameter but also possess its variation along the radius peculiar to the voids in the large scale matter distribution in the Universe \cite{K_sky1,K_sky2}; the void nature of the Cold Spot is among the discussed interpretations \cite{Masina,R}.

\begin{figure}[!htbp]
\begin{center}
\includegraphics[width=\textwidth]{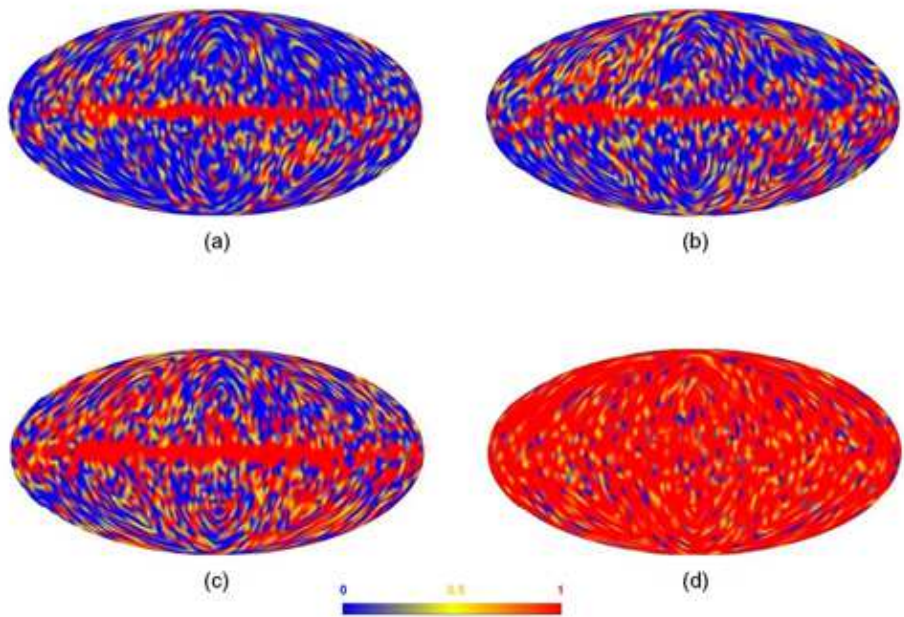} 

\caption{$\Phi(\Lambda)$ maps calculated for WMAP W,V,Q-band temperature data (a-c), respectively, and for the simulated, no-noise (d) CMB maps.}
\end{center}
	\label{fig:CMBFI}
\end{figure}

We compute the Kolmogorov stochasticity parameter and then Kolmogorov's distribution for real CMB maps, namely, for the 3 bands of the Wilkinson Microwave Anisotropy Probe's (WMAP's) temperature data in which the nature and the level of the Galactic radiations is different. We also estimate the simulated maps which all together indicate that Kolmogorov's parameter indeed
is different for maps with different contamination. This motivates the interest to model the behavior of Kolmogorov's parameter for the CMB (continuing \cite{mod1}), via specially constructed sequences composed of subsystems of different statistic, i.e. with expected statistical properties similar to the perturbations present in the primary signal in CMB maps. Such 'numerical laboratory' results show that in certain cases one can clearly distinguish the signals due to the qualitative differences in the behavior of Kolmogorov's parameter, in other cases, the quantitative properties are the indicative ones.

\section{Kolmogorov statistic}

First let us recall the definition of the Kolmogorov stochasticity parameter \cite{Kolm,Arnold}.
Consider a finite random sequence of real numbers $x_1, x_2, \ldots , x_n$, so that $x_n$ are sorted in an increasing manner  $x_1 \leq x_2 \leq \ldots \leq x_n$.

The empirical distribution function is defined as 

$$F_n(X)= (relative\, normalized\, number\, of\, the\, elements\, x_i\, which\, are\, less\, than\, X),$$
so that
\begin{eqnarray*}
F_n(X)=
\begin{cases}
0\ , & X<x_1\ ;\\
k / n\ , & x_k\le X<x_{k+1},\ \ k=1,2,\dots,n-1\ ;\\
1\ , & x_n\le X\ .
\end{cases}
\label{eq:empiricdistribution}
\end{eqnarray*}

The theoretical distribution  function $F(X)$ is defined as 

$$F(X) = (probability\, of\, the\, event\, x \leq X).$$

Kolmogorov's stochasticity parameter $\lambda_n$  for a sequence  of $n$ values of $x$ is

\begin{equation}
\lambda_n =  \sqrt{n} \cdot sup|F(X)-F_n(X)|,\\
\label{eq:lambda}
\end{equation}

so that $\lambda_n$ itself is a random variable.

According to the Kolmogorov's theorem $\lambda_n$ is a random number having empirical distribution function $$\Phi_n(\Lambda) = (relative\, normalized\, number\, of\, the\, elements\, \lambda_n\, less\, than\, \Lambda),$$ 
uniformly converging to a function $\Phi(\Lambda)$ at $n \rightarrow \infty$ for any continuous distribution $F(X)$
\begin{equation}
\Phi_n(\Lambda) \rightarrow \Phi(\Lambda),
\label{eq:philimit}
\end{equation}

where 

\begin{equation}
\Phi(\Lambda) = \sum_{k=-\infty}^{+\infty}{(-1)^k e^{-2k^2\Lambda^2}},\\
\ -\infty <  k < +\infty.
\label{eq:phi}
\end{equation}

$\Phi(\Lambda)$ is Kolmogorov's distribution, it varies from $\Phi(0)=0$ to $\Phi(\infty)=1$ monotonically.

Table 1 shows the Kolmogorov's distribution $\Phi(\Lambda)$ and its derivative $\phi(\Lambda)=\Phi'(\Lambda)$
at the values of interest of $\Lambda$.

\begin{center}
  \begin{tabular}{ | l | c | r | }
    \hline
    
    $\Lambda$ & $\Phi(\Lambda)$            & $\phi(\Lambda)$\\ \hline
    0.2       &    $ 5.05\cdot10^{-13}$    &    $1.53\cdot10^{-10}$                        \\ \hline
    0.4       &      0.0028                &     0.1012                                    \\ \hline
    0.6       &      0.1357                &     1.3241                                    \\ \hline
    0.8       &      0.4558                &     1.627                                     \\ \hline
    1         &      0.73                  &     1.071                                     \\ \hline
    1.2       &      0.8877                &     0.5385                                    \\ \hline
    1.4       &      0.9603                &     0.2222                                    \\ \hline
    1.6       &      0.988                 &     0.0764                                    \\ \hline
    1.8       &      0.996932              &     0.022                                     \\ \hline
    2         &      0.999329              &     0.0053                                    \\ \hline
    2.2       &      0.999875              &     0.0011                                    \\ \hline
    2.4       &      0.99998               &     0.00019                                   \\ \hline
  
  \end{tabular}
\end{center}

From Table 1 we see that the probable values of $\lambda$ are between 0.4 and 1.8.

According to the definition, for large enough $n$ and random sequence $x_n$ the Kolmogorov stochasticity parameter $\lambda_n$ will have a distribution close to $\Phi(\Lambda)$. If the sequence is not random, the distribution will be different. So, $\Phi(\Lambda)$ denotes the degree of randomness of a sequence.

For each studied sequence one can get only one value of $\lambda_n$. But that is not enough to conclude on the degree of randomness of that sequence. Therefore, considering sequences of large enough length $n$, we split them into subsequences, and then calculate $\lambda_n$ for each subsequence.

Therefore by our strategy we will consider the splits of the sequence $x_n$ into $m$ subsequences. In each case we deal with new sequences $\lambda_n^m$ of length $m$, which should be a random sequence of a distribution close to $\Phi(\Lambda)$ according to Kolmogorov's theorem.

\section{CMB $\Phi$ maps}

We computed the Kolmogorov stochasticity parameter $\lambda_n$ for the CMB temperature maps, then obtained the Kolmogorov's distribution $\Phi$ over the sky. CMB temperature maps are given in HEALPIX coordinate system that can have different resolutions described by parameter $NSide$. We have used WMAP's  CMB maps of $NSide=512$,  in order to have enough temperature sequences to obtain lower resolution $NSide=32$ $\Phi$ maps, i.e. we reconstruct a sequence from pixels in one cell of lower resolution map and calculate $\Lambda$ then $\Phi$; for more details see \cite{K_sky1,K_sky2}. We used the WMAP's 5-year data \cite{H} in the W,V and Q bands (Fig.1 a-c). It is known that the contamination of the cosmological CMB signal is different in these maps due to the different contribution mainly of the Galactic synchrotron and the dust radiation (for details see \cite{H,DL} and references therein). 

As it seen from Figure 1a-c the $\Phi$-maps are different for different bands, which shows that Kolmogorov's distribution is indeed sensitive to the degree of the contamination by specific signals. Then, in Fig.1d we show the $\Phi$-map for the temperature map simulated by a standard scheme for the power spectrum parameters, i.e. the purely cosmological signal, without any noise. The difference of the simulated $\Phi$-map from those of real maps is also visible. This indicates the principal possibility of the simulation of a given signal or several signals and then the comparison of their $\Phi$-maps with the observed ones, in order to reveal the contribution of a given component. However, to do that one has to model signals with various compositions of sub-signals, to probe their sensitivity to the Kolmogorov's statistic.

\section{The sequences and their properties}

We have analyzed sequences of uniform distribution: 7000 different sequences of length 10000 each were generated, which then were divided into 7 groups, each of 1000 sequences.

The list of the considered sequences is as follows: 

I: random;

II:  non-random: 
\begin{equation}
x_n=107\, n\, mod\, 513;
\end{equation}

III: sequences including two copies of the same random sequence, 
\begin{equation}
x_1, x_2, \ldots , x_{n/2},x_1, x_2, \ldots , x_{n/2};
\end{equation}

IV: four copies of a same random sequence;

V: random with perturbations, 
\begin{equation}
x_i=\xi_i+\epsilon sin{\xi_i},
\end{equation} 
with parameter $\epsilon=0.03$;

VI: the same as V but with $\epsilon=0.01$;

VII: non-random, where the two neighbors are equal: 
\begin{equation}
x_1, x_1,x_2, x_2, \ldots , x_n, x_n.
\end{equation}


\begin{figure}[h]
\begin{minipage}[h]{0.49\linewidth}
\center{\includegraphics[width=1.0\linewidth]{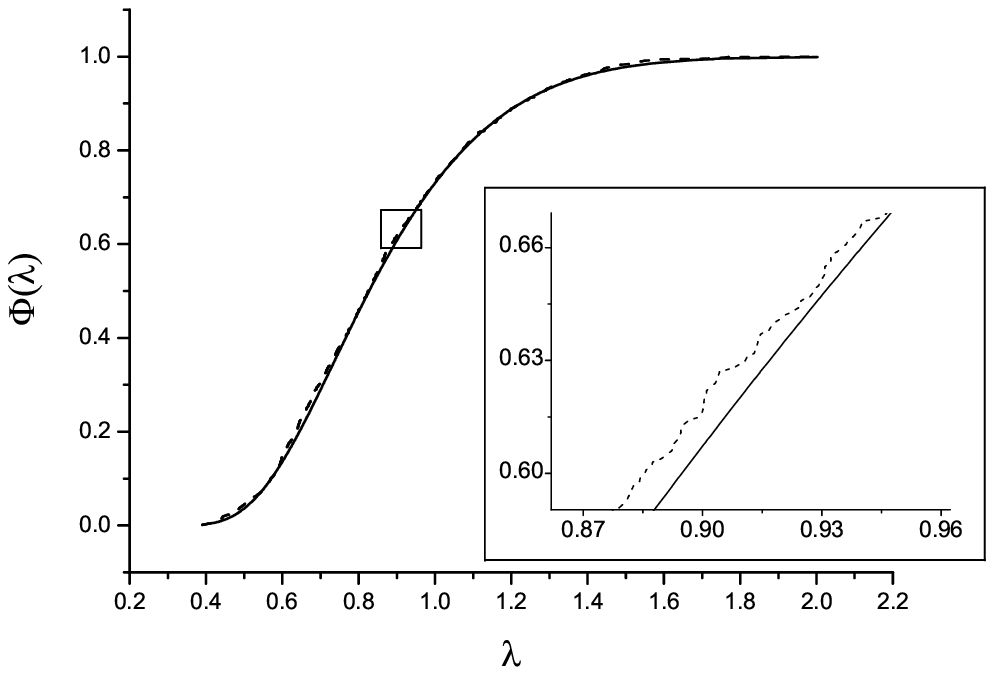} \\ (a)}
\end{minipage}
\hfill
\begin{minipage}[h]{0.49\linewidth}
\center{\includegraphics[width=1.0\linewidth]{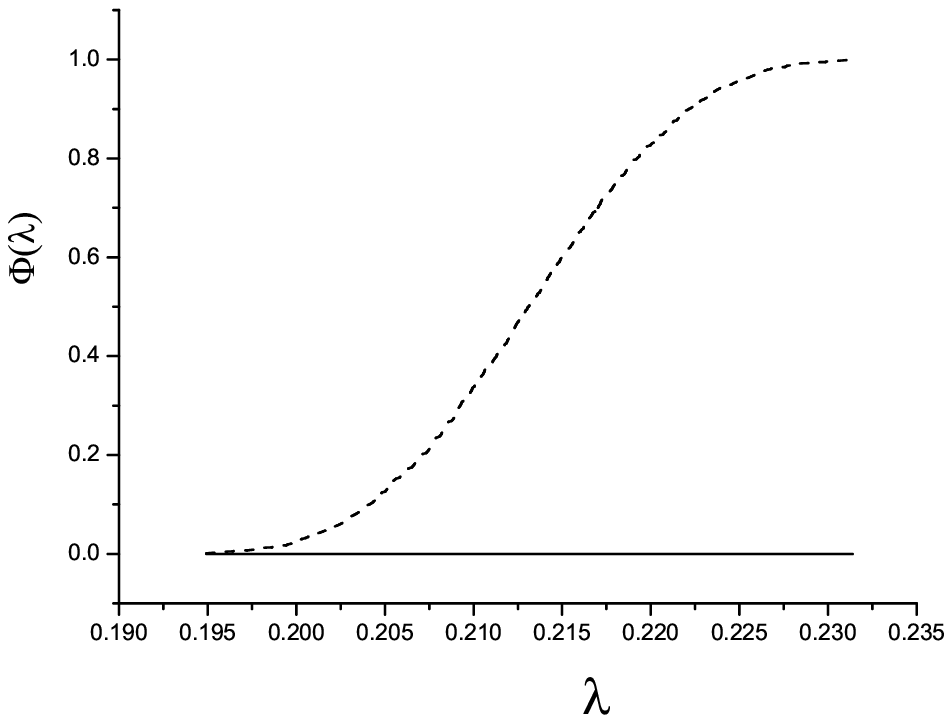} \\ (b)}
\end{minipage}
\vfill
\begin{minipage}[h]{0.49\linewidth}
\center{\includegraphics[width=1.0\linewidth]{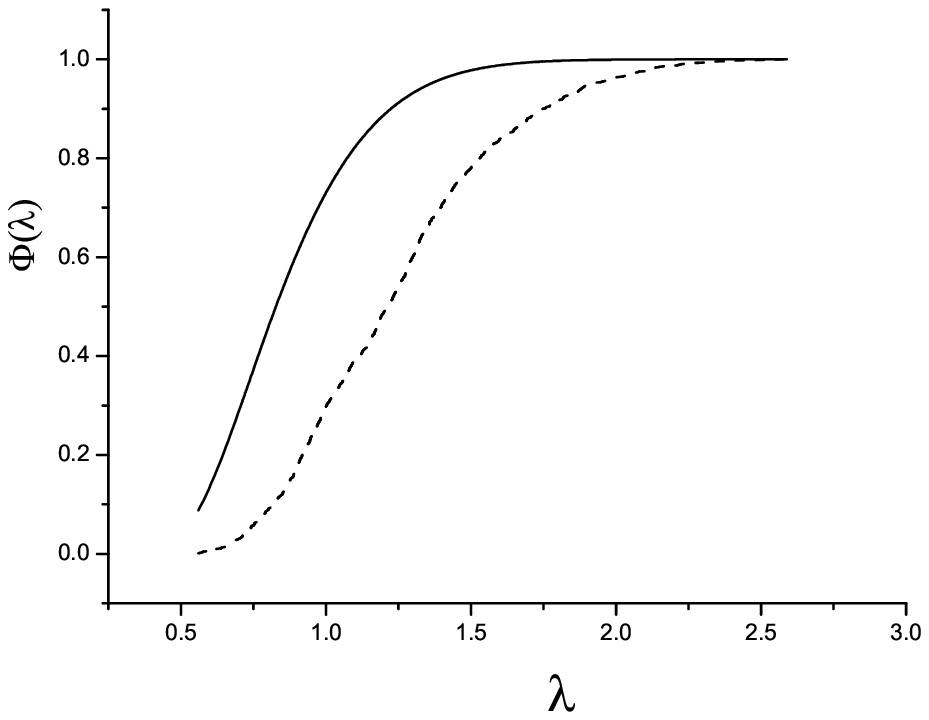} \\ (c)}
\end{minipage}
\hfill
\begin{minipage}[h]{0.49\linewidth}
\center{\includegraphics[width=1.0\linewidth]{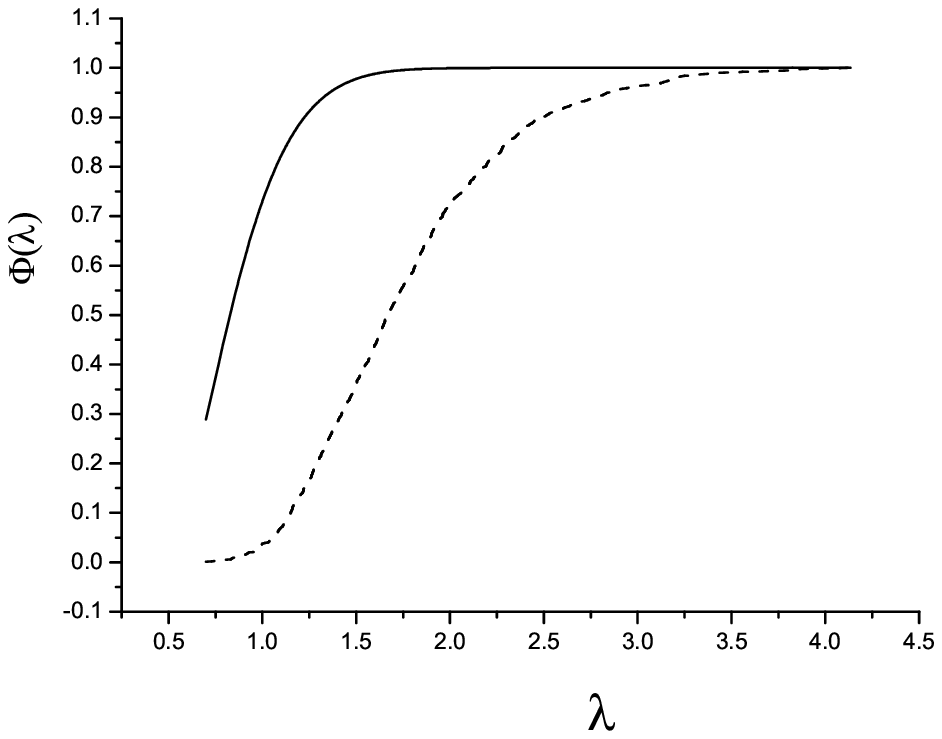} \\ (d)}
\end{minipage}
\vfill
\begin{minipage}[h]{0.49\linewidth}
\center{\includegraphics[width=1.0\linewidth]{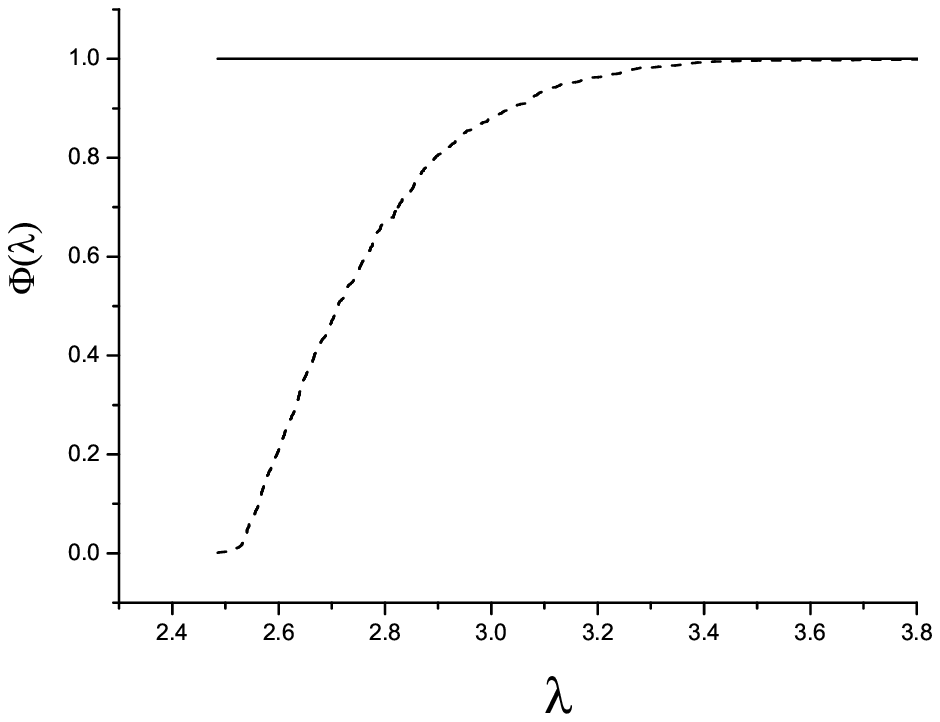} \\ (e)}
\end{minipage}
\hfill
\begin{minipage}[h]{0.49\linewidth}
\center{\includegraphics[width=1.0\linewidth]{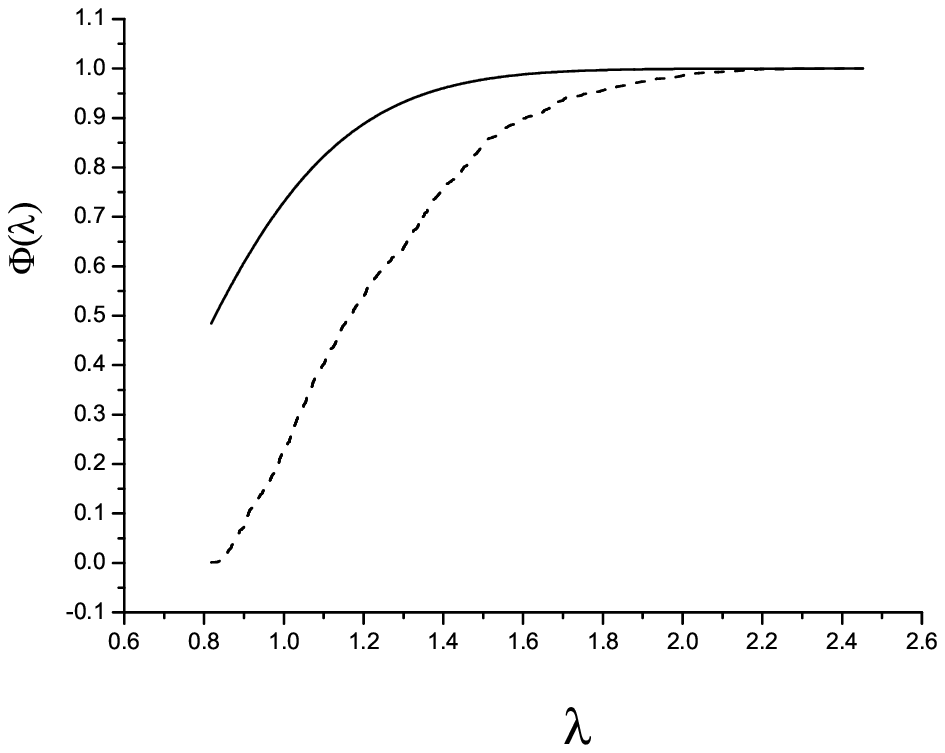} \\ (f)}
\end{minipage}
\vfill
\begin{minipage}[h]{0.49\linewidth}
\center{\includegraphics[width=1.0\linewidth]{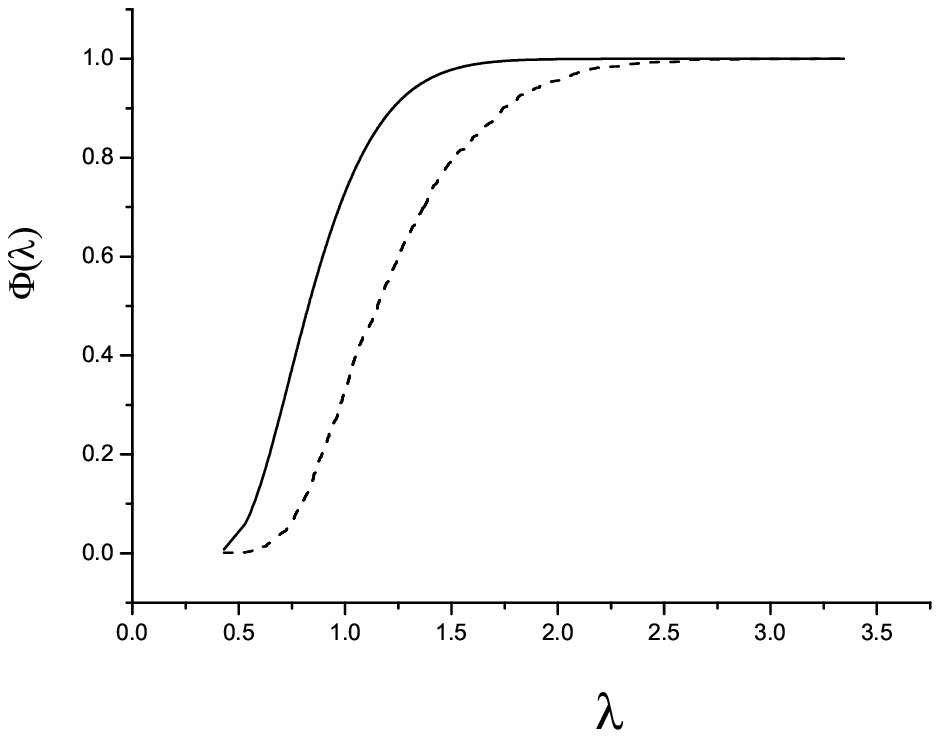} \\ (g)}
\end{minipage}
\caption{ Kolmogorov's distribution $\Phi(\lambda)$ (solid line) and empirical distribution $\Phi_n(\lambda)$ (dashed line) for calculated values of $\lambda_n$ for the considered groups: (a) I, (b) II, (c) III, (d) IV, (e) V, (f) VI and (g) VII.}
	\label{fig:Graph}
\end{figure}

Figure \ref{fig:Graph} shows the results of the computations. In all figures the horizontal axis is $\lambda$, the vertical axis denotes $\Phi(\lambda)$, the dashed line is the empirical distribution function, the solid line is the $\Phi(\lambda)$.   The figures shows the level of coincidence of  $\Phi(\lambda)$ with the empirical distribution function for each group.

We see that only for random sequences $\Phi(\lambda)$ practically coincide with the empirical distribution function, as it is expected by the Kolmogorov's theorem. In all other cases the $\Phi(\lambda)$ deviates from the empirical distribution function. For example, for non-random sequences of type $x_n=107$ $n$ mod $513$ $\Phi(\lambda)$ is quite different from the empirical distribution function, as the values of $\lambda$ vary between 0.195 and 0.230. For sequences including two and four copies of the same random sequence, the values of $\lambda$ are, respectively, $\sqrt{2}$ and 2 times larger than for those for the random case due to the factor $\sqrt{n}$ in the expression for $\lambda$. In the perturbed case with $\epsilon=0.03$ we have much bigger values of $\lambda$, so $\Phi(\lambda)=1$.
 
Table 2 contains  $\Phi(\lambda)$ and $\Phi_n(\lambda)$ for the I group, the random sequences.

 \begin{center}
  \begin{tabular}{ | l | c | r | }
    \hline
    $\lambda$ & $\Phi(\lambda)$ & $\Phi_n(\lambda)$ \\ \hline
    0,38996	  & 0,00193         &	1E-3              \\ \hline
    0,5707	  & 0,09945	        & 0,101             \\ \hline
    0,63823   &	0,19            &	0,201             \\ \hline
    0,69643   &	0,28284         & 0,301             \\ \hline
    0,76231	  & 0,39352         &	0,401             \\ \hline
    0,82529	  & 0,4964          &	0,501             \\ \hline
    0,88579	  & 0,58735	        & 0,601             \\ \hline
    0,97301   &	0,69994	        & 0,701             \\ \hline
    1,07263	  & 0,7999	        & 0,801             \\ \hline
    1,23134   & 0,90361	        & 0,901             \\ \hline
 \end{tabular}
\end{center}

 Then we split the sequences into 50 subsequences, obtain 50 values of $\lambda$ for each subsequence, and calculate $\lambda_{mean}$. We are interested in the difference between the values of $\lambda_{mean}$ and theoretical mean $\lambda_{Mean}=\int{\lambda \phi(\lambda) d\lambda}\approx 0.875029$ where $\phi(\lambda)=\Phi'(\lambda)$. For random sequences  $\lambda_{mean}$ must be equal to theoretical mean value $\lambda_{Mean}\approx 0.875029$. Therefore this difference is a
 direct measure that describes the degree of the randomness of a sequence, bigger is that difference, more regular is the sequence, for details see \cite{Arnold,Arnold1}.

\begin{figure}[!htbp]
	\centering
		\includegraphics[width=4 in]{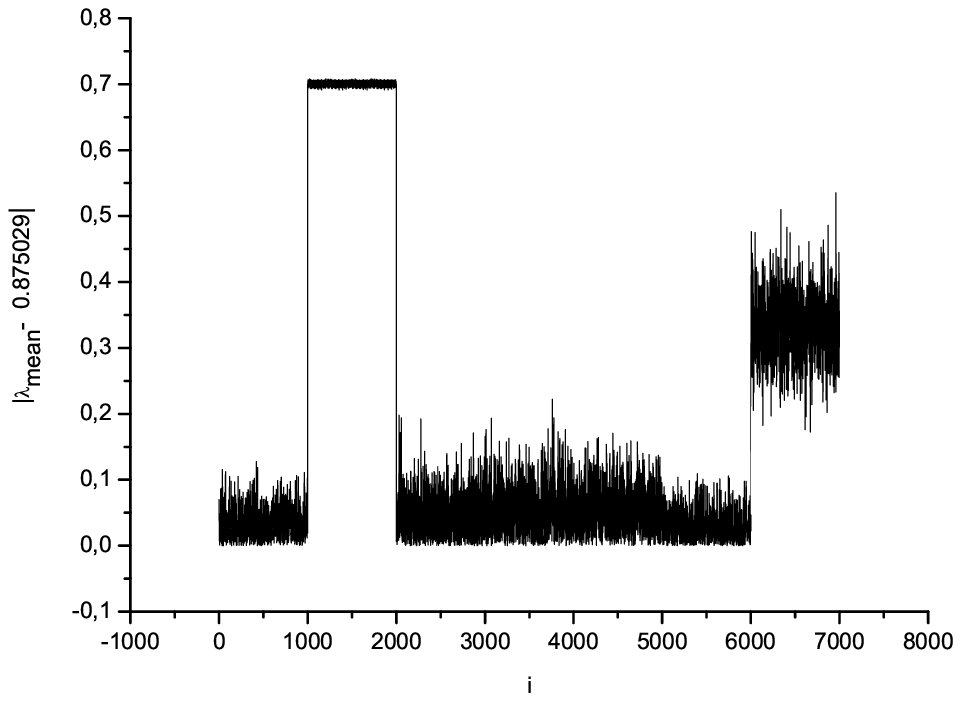}
		\caption{Normalized $\lambda_{mean}$ for all sequence groups, $i$ denoting the number of the sequence.}
	\label{fig:lambdamean}
\end{figure}

\begin{figure}[!htbp]
	\centering
		\includegraphics[width=4 in]{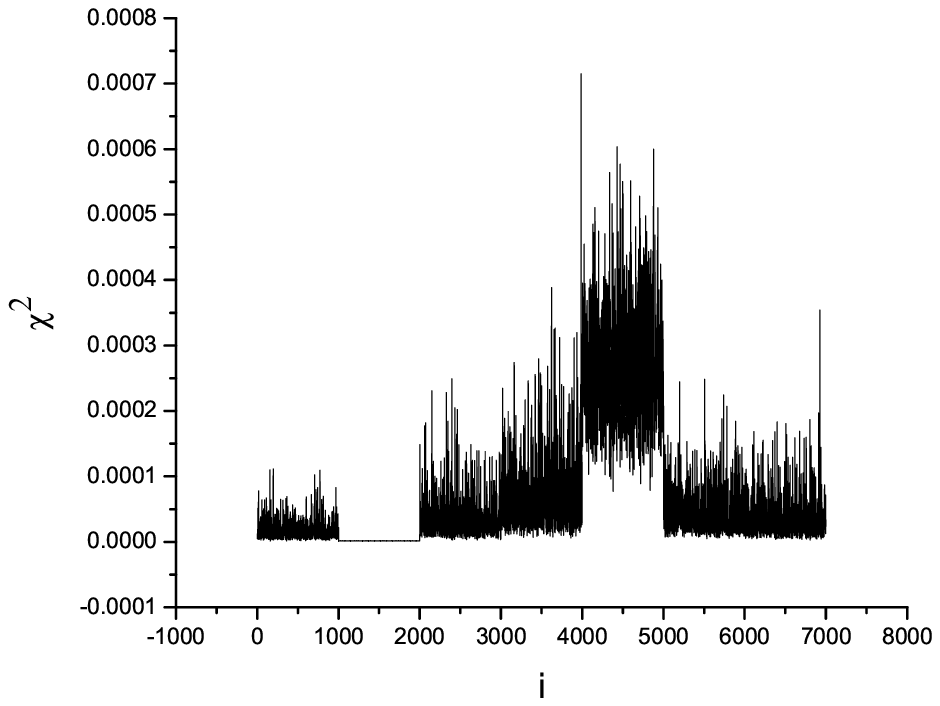}
		\caption{The same as in Fig. 2 but for $\chi^2$.}
	\label{fig:chi2}
\end{figure}

In Figures \ref{fig:lambdamean} and \ref{fig:chi2} horizontal axis $i$ is the index of the sequence, so each 1000 points corresponds to one group noted above. Figure \ref{fig:lambdamean} shows that, the difference between groups I and III or IV are not obvious, i.e. the split subsequences do not "feel" the duplicates. Also, for V and VI groups the perturbations are small, so again there is no visible difference between them and the group I. The random case differs from II and VII groups, because of the different scales.

Then we found out the behavior of $\chi^2$ for each group of sequences. $\chi^2$ is calculated for $F(X)$ and $F_n(X)$ and shows how far is the distribution from the uniform one. Here $\chi^2$ is defined as

\begin{equation}
\chi^2=\sum\limits_{i=1}^n\frac{(F_n(x_i)-F(x_i))^2}{n}. 
\end{equation}

\begin{figure}[h]
\begin{minipage}[h]{0.49\linewidth}
\center{\includegraphics[width=1.0\linewidth]{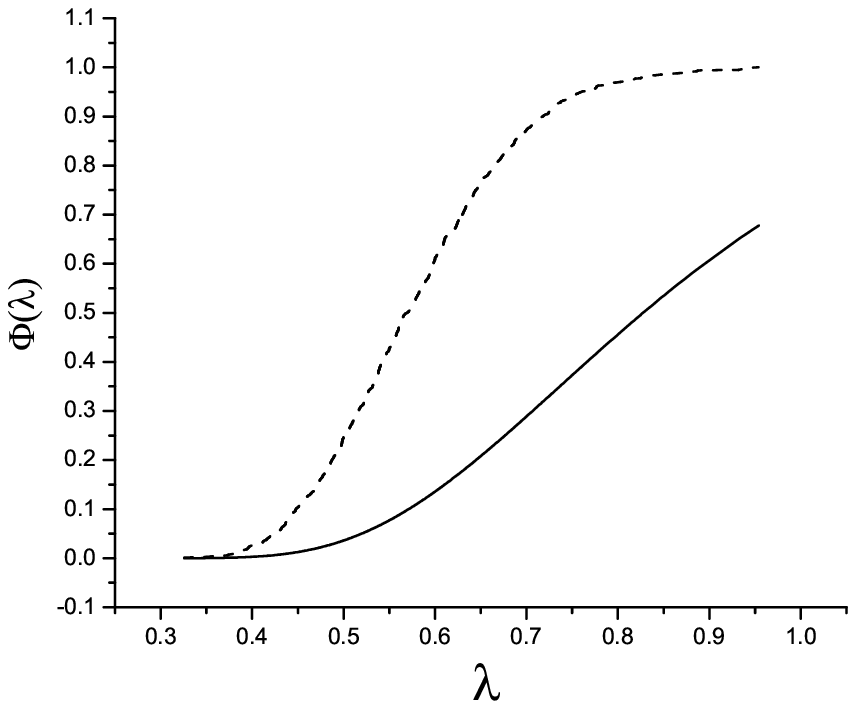} \\ (1)}
\end{minipage}
\hfill
\begin{minipage}[h]{0.49\linewidth}
\center{\includegraphics[width=1.0\linewidth]{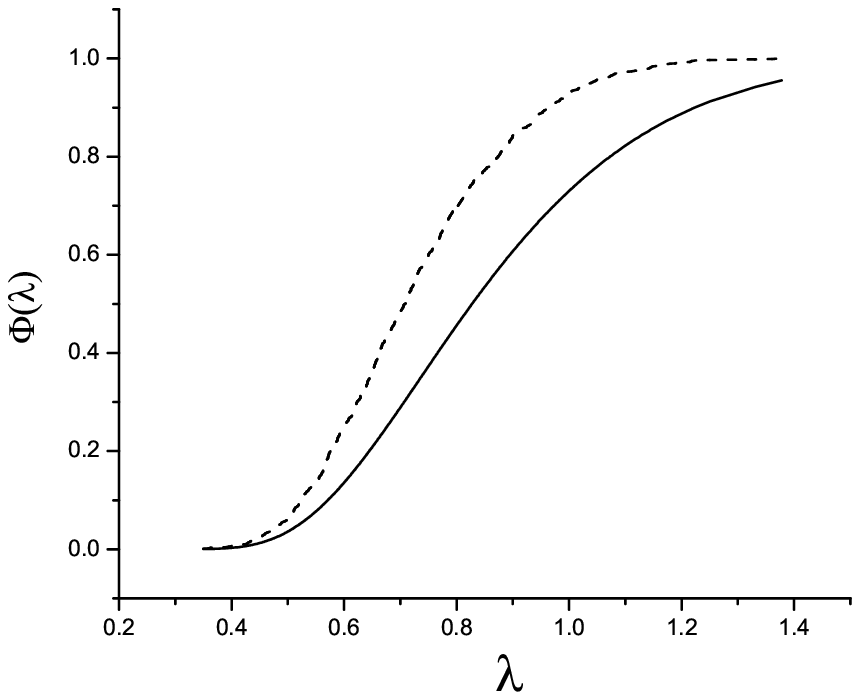} \\ (2)}
\end{minipage}
\vfill
\begin{minipage}[h]{0.49\linewidth}
\center{\includegraphics[width=1.0\linewidth]{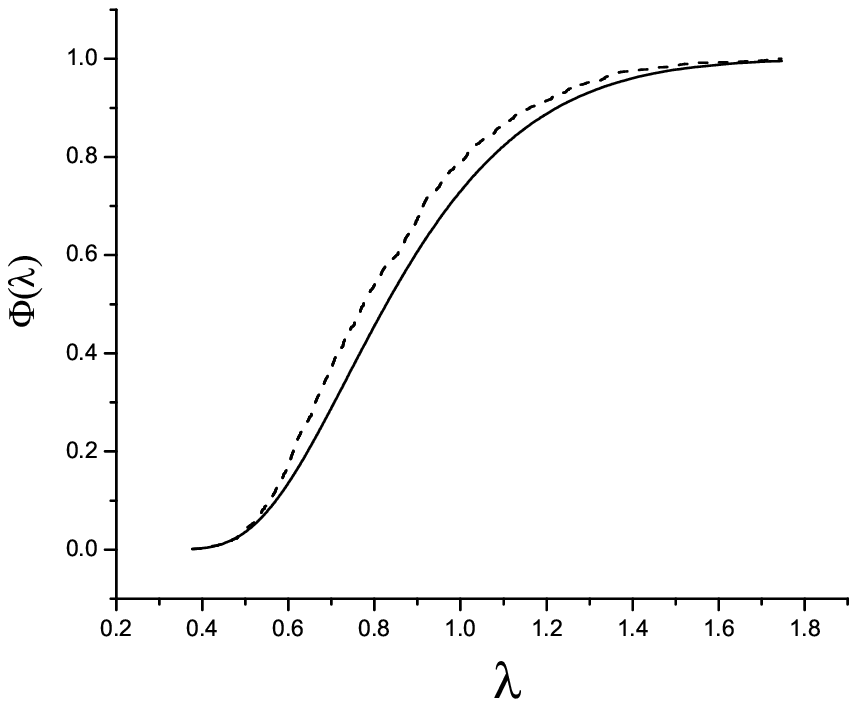} \\ (3)}
\end{minipage}
\hfill
\begin{minipage}[h]{0.49\linewidth}
\center{\includegraphics[width=1.0\linewidth]{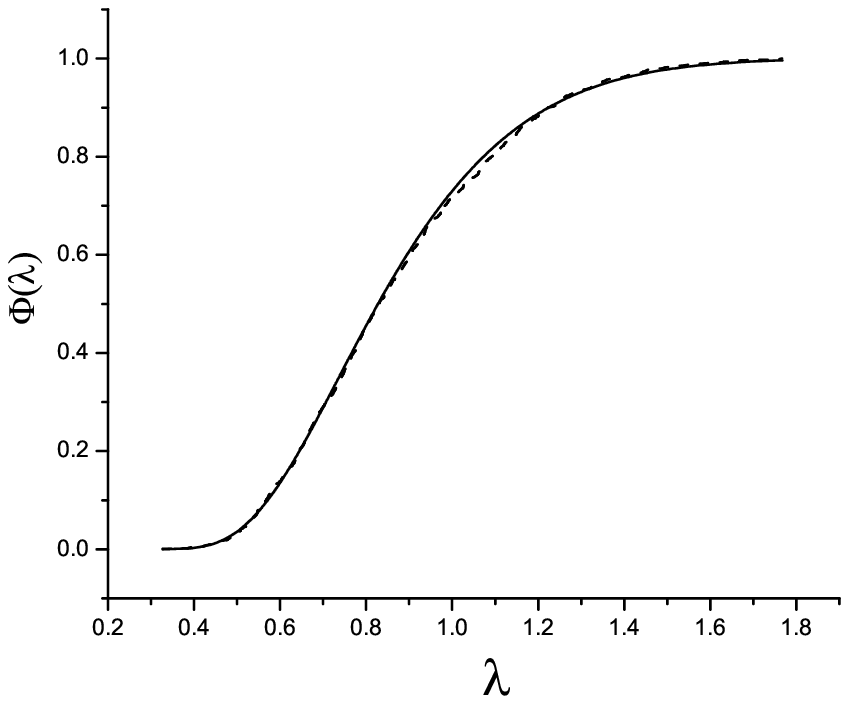} \\ (4)}
\end{minipage}
\caption{ Kolmogorov's distribution $\Phi(\lambda)$ for $z_n=\alpha x_n+(1-\alpha)y_n$ for (1) $\alpha=0.2$, (2) $\alpha=0.4$, (3) $\alpha=0.6$, (4) $\alpha=0.8$. As in Fig. 3, the solid line denotes the Kolmogorov distribution (3), the dashed line denotes the computed empirical distribution.}
	\label{fig:a}
\end{figure}

To generalize the considered above results, we studied sequences of the type 

\begin{equation}
z_n=\alpha x_n+(1-\alpha)y_n, 
\label{form:zn}
\end{equation}
where $\alpha$ between $1$ and $0$ specifies the limiting the random $x_n$, and regular of the form $y_n=107i\, mod\, 513$, cases, respectively. 101 sequences were studied, each containing 100 subsequences of length 10000. Steps of 0.01 were used for $\alpha$, so, to each $\alpha$ correspond 100 sequences and hence we have 10100 sequences. Plots of $\Phi(\Lambda)$ vs $\Lambda$ are shown for different values of $\alpha$ in Figure \ref{fig:a}.

\begin{figure}[h]
\begin{minipage}[h]{0.49\linewidth}
\center{\includegraphics[width=1.0\linewidth]{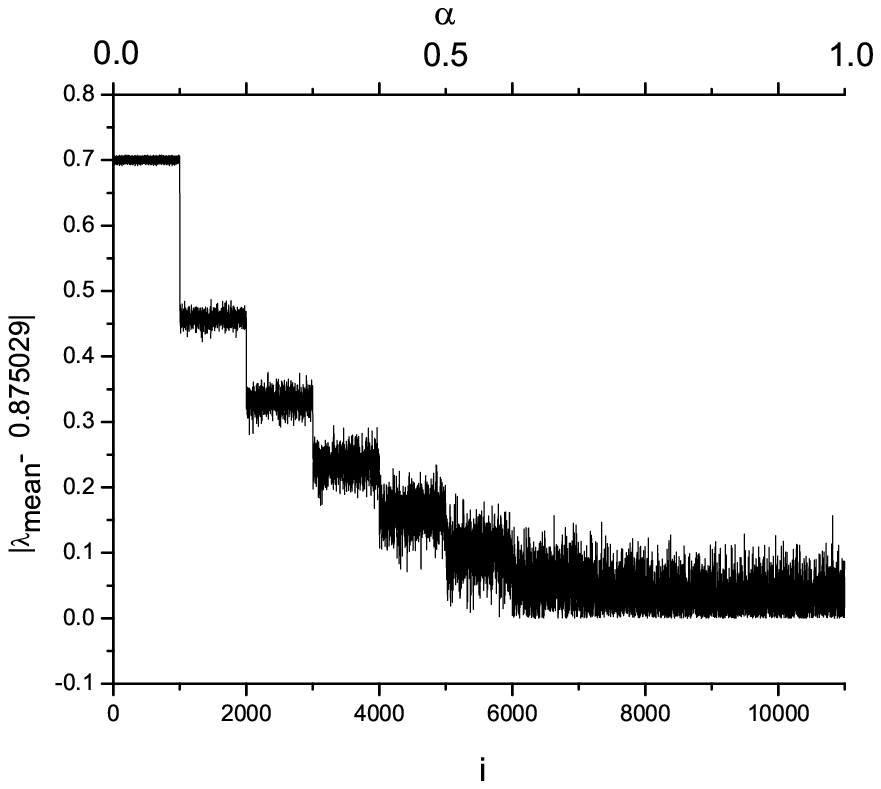} \\ (a)}
\end{minipage}
\hfill
\begin{minipage}[h]{0.49\linewidth}
\center{\includegraphics[width=1.0\linewidth]{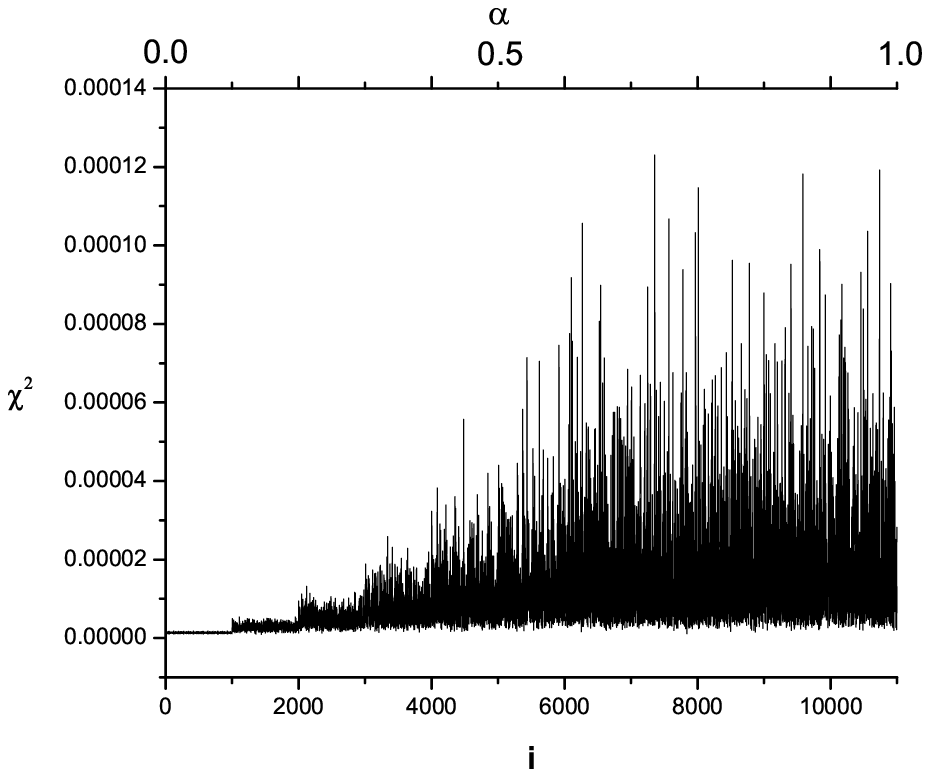} \\ (b)}
\end{minipage}
 \caption{(a) Normalized $\lambda_{mean}$ for all $\alpha$, (b) $\
$ for all $\alpha$. $i$ is the number of sequences. each 1000 points correspond to one value of $\alpha$ which increases with step $0.1$.}
\label{ris:image1}
\end{figure}

From the Figure \ref{ris:image1}(a) we can see that when $\alpha=0$ (the regular case), the normalized $\lambda_{mean}$ precisely corresponds to the normalized value for the second group of Figure \ref{fig:lambdamean}. For $\alpha=1$ (the random case), the normalized $\lambda_{mean}$ precisely corresponds to the normalized $\lambda_{mean}$ for the second group shown in Figure \ref{fig:lambdamean}. 

In Figure \ref{ris:image1} $\chi^2$ and normalized $\lambda_{mean}$ for $\alpha=0$ and $\alpha=1$ cases behave the same way as in 2nd and 1st groups respectively in figures \ref{fig:chi2} and \ref{fig:lambdamean}.

We were also interested in the behavior of $\lambda_{mean}$ depending on the parameters of sequences i.e. the number of sequences, their length, number of the splits, type of $y_n$, as well as on the steps of $\alpha$.

\begin{figure}[h]
\begin{minipage}[h]{0.49\linewidth}
\center{\includegraphics[width=1.0\linewidth]{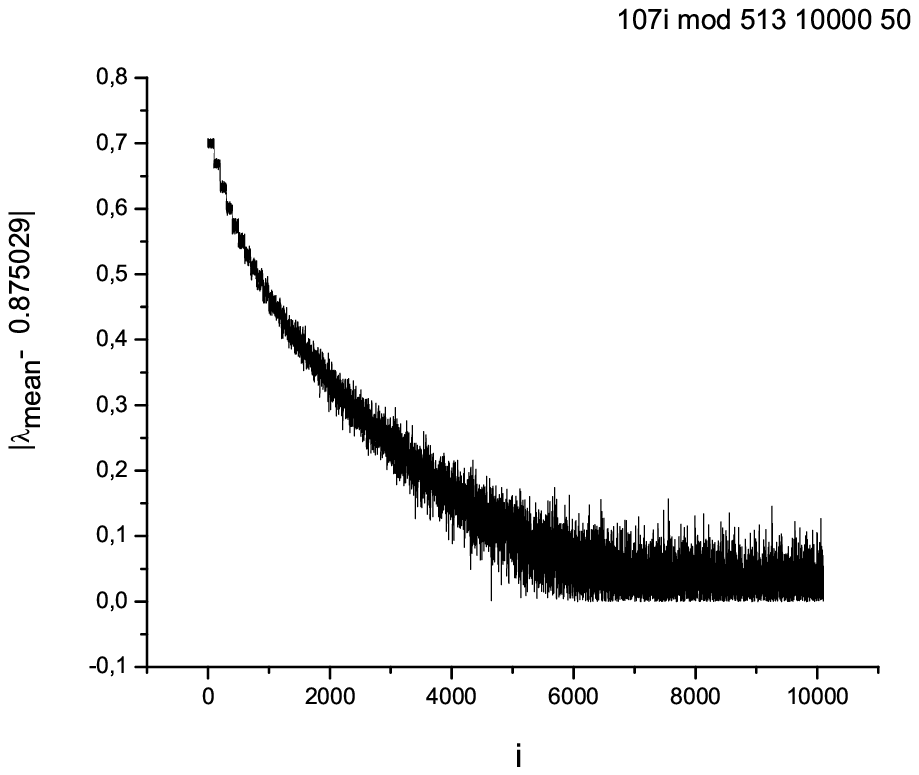} \\ (a)}
\end{minipage}
\hfill
\begin{minipage}[h]{0.49\linewidth}
\center{\includegraphics[width=1.0\linewidth]{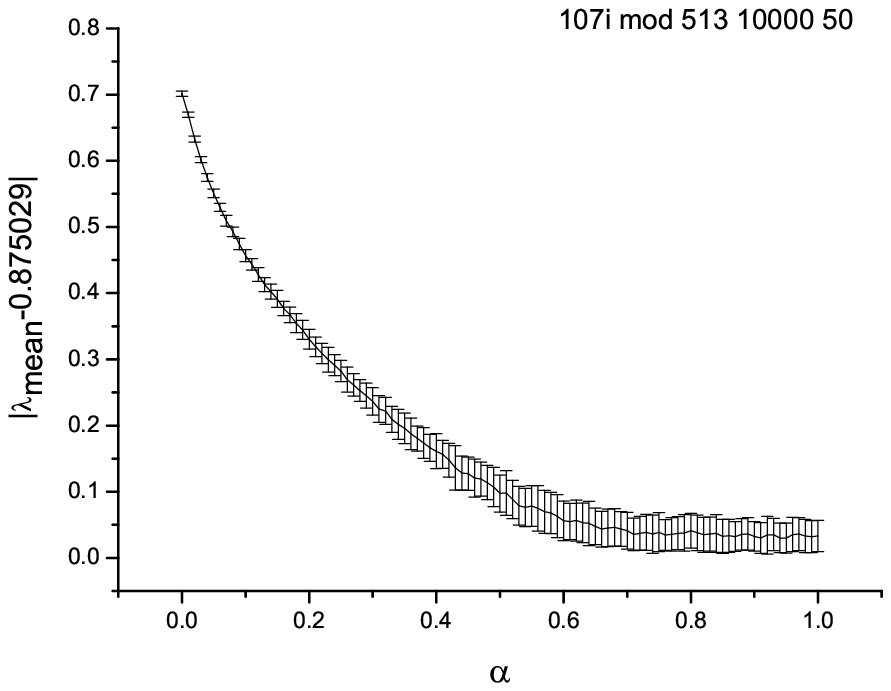} \\ (b)}
\end{minipage}
\vfill
\begin{minipage}[h]{0.49\linewidth}
\center{\includegraphics[width=1.0\linewidth]{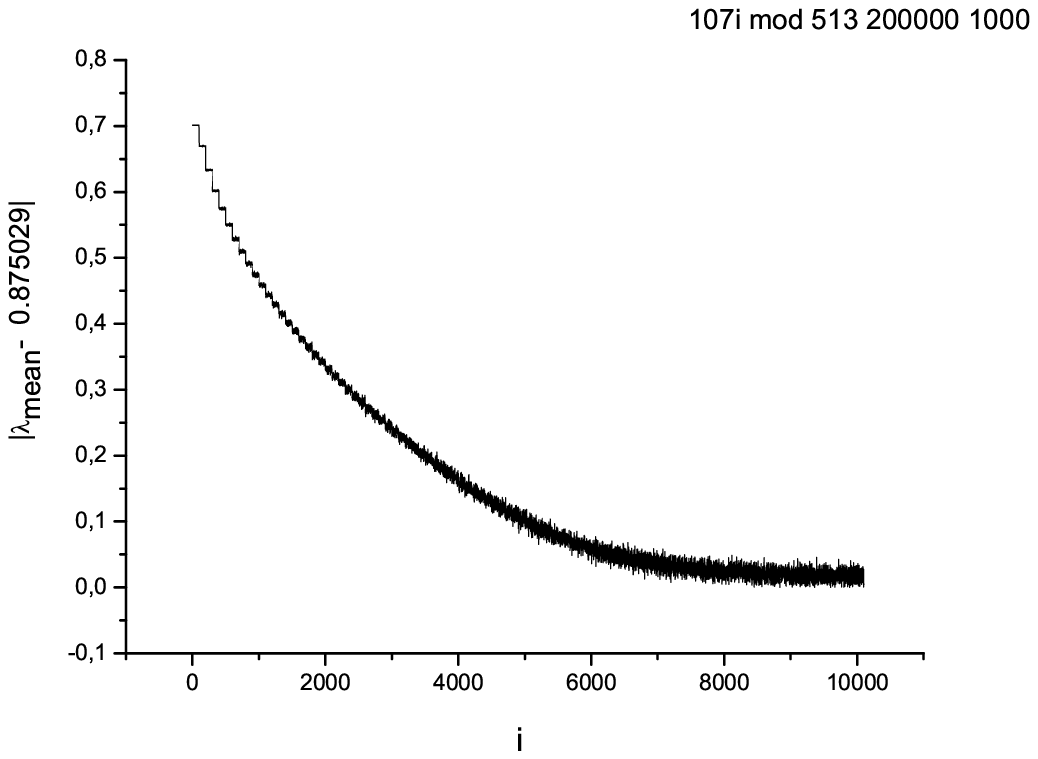} \\ (c)}
\end{minipage}
\hfill
\begin{minipage}[h]{0.49\linewidth}
\center{\includegraphics[width=1.0\linewidth]{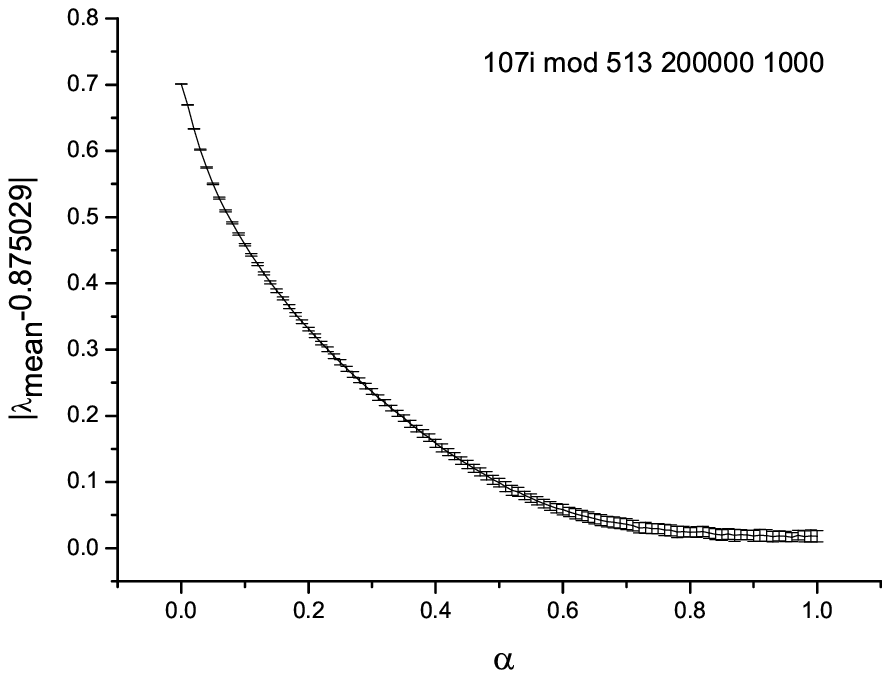} \\ (d)}
\end{minipage}
\caption{The behaviour of normalized $\lambda_{mean}$ depending on value of $\alpha$ for $z_n$ in (\ref{form:zn}); (b,d)
are the smoothed ones of (a,c). The form of $y_n$, the number of sequences and of their subgroups are given above the curves.}
\label{fig:exp}
\end{figure}

We also changed the type of $y_n$, their length and the number of splits, then fitted the corresponding numerical results with an exponential decaying function. Figure \ref{fig:exp} shows the results of computations. The left plots are the normalized $\lambda_{mean}$ dependences on the index of the sequence, the right plots are the corresponding fits.
In each plot the legend above indicates the type of the generated sequences, the length and number of splits: for example, the Figure \ref{fig:exp}(a) is a result of the generated sequences of the type $107i\, mod\, 513$, 10000 is the number of elements in each sequence and 50 is the number of splits of the sequences.

One can see that the maximum of 
\begin{equation}
|\lambda_{mean}-0.875029|
\end{equation} 
depends on the parameters of $y_n$, the length of sequences and the number of splits, however, in the cases where the length and the number of splits ratios are equal, maximal values of $|\lambda_{mean}-0.875029|$ are also equal to a precision about $10^{-3}$.

When larger length sequences were applied, the error bars of the belt in Figure \ref{fig:exp} decrease and due to the large numbers of points the value of $\chi^2$ increases, as shown in the legends.

\section{Conclusions}

We analyzed several types of sequences using the Kolmogorov stochasticity parameter to model the contributions of signals of different statistic in the Cosmic Microwave Background radiation. The previously undertaken estimations of the Kolmogorov's parameter for the real CMB temperature maps obtained by WMAP had revealed the potential of the method in the separation of the Galactic disk and of other structures in the previous works \cite{K_sky1,K_sky2}. We have also computed the Kolmogorov CMB maps for various frequency bands of WMAP, which show the variation of the statistic at various degree of contamination of the signal by the Galaxy. Those peculiarities in the properties of the CMB temperature anisotropies have defined the strategy of our present study. Namely we did as follows: once we saw that the Galactic contribution can be separated in CMB maps with Kolmogorov analysis due to its different degree of randomness, we modeled systems comprised of signals that differ in their randomness. Such strategy in principle can help to reveal the role of a given sub-signal.  

Although our numerical experiments deal with specific sequences, the performed analysis shows that the CMB can be in principle be separated into sub-signals using the Kolmogorov statistic and the stochasticity parameter, either qualitatively or quantitatively. 

Some of the systems considered above can serve as numerical illustrations to the examples discussed in \cite{Arnold,Arnold1}.
 
We are grateful to V.G. Gurzadyan, A.A. Kocharyan, T. Ghahramanyan and G. Yegorian for discussions and help and to the referee for many valuable comments.

\end{document}